\documentclass[12pt]{article}
\usepackage{cite}
\usepackage{CJK}
\usepackage{amssymb}
\usepackage{amsmath}
\usepackage{enumerate}
\usepackage{graphicx}
\usepackage{amsfonts}
\usepackage{url}
\usepackage{bm}

\usepackage{pifont}
\usepackage{epsfig,subfigure,dsfont,amsthm,amsbsy,mathrsfs,amscd}

\def\la{\langle}
\def\ra{\rangle}
\def\be{\begin{equation}}
\def\ee{\end{equation}}
\def\ba{\begin{array}}
\def\ea{\end{array}}
\input amssym.def

\newcommand\btd{\raise 2pt \hbox{$\hat\bigtriangledown$}\hskip 1.5pt}
\newcommand\bt{\raise 2pt \hbox{$\bigtriangledown$}\hskip 1.5pt}
\begin{document}
 \title{\large\bf Quantum Nonlocality of Arbitrary Dimensional Bipartite States}
\author{Ming Li$^\dag$$^\flat$, Tinggui Zhang$^\ddag$$^\flat$, Bobo Hua$^\S$$^\flat$, Shao-Ming
Fei$^\sharp$$^\flat$ and Xianqing Li-Jost$^\flat$$^\ddag$\\[10pt]
\footnotesize \small $^\dag$College of the Science, China University
of Petroleum,\\
\small Qingdao 266580, P. R. China\\
\footnotesize
\small $^\ddag$College of Mathematics and Statistics, Hainan Normal University,\\
\small Haikou 571158, P. R. China\\
\small $^\S$School of Mathematical Sciences, LMNS, Fudan University,\\
\small Shanghai 200433, P. R. China\\
\small $^\sharp$School of Mathematical Sciences, Capital Normal University,\\
\small Beijing 100048, P. R. China\\
\small $^\flat$Max-Planck-Institute for Mathematics in the Sciences,\\
\small Leipzig 04103, Germany}
\date{}

\maketitle

\centerline{$^\ast$ Correspondence to tinggui333@163.com}
\bigskip

\begin{abstract}

We study the nonlocality of arbitrary dimensional bipartite quantum
states. By computing the maximal violation of a set of multi-setting
Bell inequalities, an analytical and computable lower bound has been
derived for general two-qubit states. This bound gives the necessary
condition that a two-qubit state admits no local hidden variable
models. The lower bound is shown to be better than that from the
CHSH inequality in judging the nonlocality of some quantum states.
The results are generalized to the case of high dimensional quantum
states, and a sufficient condition for detecting the non-locality
has been presented.
\end{abstract}
\bigskip

Quantum mechanics is inherently nonlocal, as revealed by the
violation of Bell inequality \cite{bell}. A bipartite quantum state
may violates some Bell inequalities such that the local measurement
outcomes can not be modeled by classical random distributions over
probability spaces. Namely, the state admits no local hidden
variable (LHV) model.

The nonlocality and quantum entanglement play important roles in our
fundamental understandings of physical world as well as in various
novel quantum informational tasks \cite{nielsen,di}. A quantum state
without entanglement must admit LHV models
\cite{vbell1,vbell2,vbell3,chenjingling,liprl,yu}. However, not all
the entangled quantum states are of nonlocality
\cite{werner,barrett,062105,032112}. To show that a quantum state
admits a LHV model, it is sufficient to construct such LHV model
explicitly \cite{werner,062105}. To show that a quantum state admits
no LHV models, it is sufficient to show that it violates a Bell
inequality. Quantum states that violate Bell inequalities are also
useful in building quantum protocols to decrease communication
complexity \cite{dcc} and provide secure quantum communication
\cite{scc11,scc12}. Moreover, since the nonlocality is detected by
the violation of Bell inequalities, quantum nonlocality could be
quantified in terms of the maximal violation value for all Bell
inequalities. However, it is a formidable task  either to show that
a state admits an LHV model, or to show that a state violates a Bell
inequality.

Let $A_i$ and $B_i$, $i=1,2,\cdots, n$, be observables with respect
to the two subsystems of a bipartite state, with eigenvalues $\pm
1$. Let $M$ be a real matrix with entries $M_{ij}$ such that
$\max_{a_i,b_j=\pm 1}|\sum_{i,j=1}^{n}M_{ij}\,a_ib_j|=1$. Denote
$I=\sum_{i,j=1}^{n}M_{ij}A_i\otimes B_j$ the corresponding Bell
operator. Define \be\label{q} Q=\sup_{M}\max_{A_i,B_j}|\la I
\ra_{\rho}|, \ee where $\la I \ra_{\rho}=tr(I\rho)$ stands for the
mean value of the Bell operator associated to state $\rho$.
Obviously a quantum state $\rho$ can never be described by a LHV
model if and only if $Q$ is strictly larger than $1$.

In \cite{werner,barrett,062105,032112,hua}, the authors have
investigated the nonlocality of Werner states. For two-qubit Werner
state $\rho_w=x|\psi^-\ra\la\psi^-|+(1-x)\frac{I}{4}$,
$|\psi^-\ra=(|01\ra-|10\ra)/\sqrt{2}$, the quantity $Q$ is proved to
be $\frac{x}{4}K_G(3)$ in \cite{062105}, where $K_G(3)$ is the
Grothendieck's constant of order three. However, since up to now one
does not kown the exact value of the Grothendieck's constant
$K_G(3)$, $Q$ is still is not known. The upper and lower bounds of
the threshold value of this parameter $Q$ have been refined by
constructing better LHV models \cite{werner,barrett,062105} or by
finding better Bell inequalities \cite{032112,hua}.

In the paper we study the nonlocality of arbitrary two-qubit states
and present an analytical and computable lower bound of the quantity
$Q$ by computing the maximal violation of a set of multi-setting
Bell inequalities. The lower bound is shown to be better than that
derived in terms of the CHSH inequality for some quantum states. We
also present a sufficient condition that a high dimensional quantum
state admits LHV models.

\medskip
\noindent{\bf Results}
\medskip

{\sf Lower bound of $Q$ for two-qubit quantum states}~ A two-qubit
quantum state $\rho$ can be always expressed in terms of Pauli
matrices $\sigma_i$, $i=1,2,3$, \be\label{rho} \rho=\frac{1}{4}
I\otimes I +\sum_{i=1}^3 r_i\sigma_i\otimes I+\sum_{j=1}^3
s_jI\otimes\sigma_j +\sum_{i,j=1}^3 t_{ij}\sigma_i\otimes \sigma_j,
\ee where $r_k=\frac{1}{4}Tr(\rho\sigma_k\otimes I)$,
$s_l=\frac{1}{4}Tr(\rho I\otimes\sigma_l)$ and
$t_{kl}=\frac{1}{4}Tr(\rho\sigma_k\otimes\sigma_l)$. We denote $T$
the matrix with entries $t_{ij}$.

The key point in computing $Q$ is to find
$\max_{\vec{a}_i\vec{b}_j}\la I \ra$ over all $M$ under the
condition $\max_{a_i,b_j=\pm 1}|\sum_{i,j=1}^{n}M_{ij}\,a_ib_j|=1$.
In \cite{032112} a Bell operator has been introduced, \be\label{bo}
I=\frac{1}{n^2}[\sum_{i,j=1}^{n}A_i\otimes B_j+\sum_{1\leq i < j\leq
n}C_{ij}\otimes(B_i-B_j)+\sum_{1\leq i < j\leq n}(A_i-A_j)\otimes
D_{ij}], \ee where $A_i, B_j, C_{ij}$ and $D_{ij}$ are observables
of the form $\sum_{\alpha=1}^3 x_{\alpha}\sigma_{\alpha}$ with
$\vec{x}=(x_{1},x_{2},x_{3})$ the unit vectors.

To find an analytical lower bound of $Q$, we consider infinite many
measurements settings, $n\to\infty$. Then the discrete summation in
(\ref{bo}) is transformed into an integral of the spherical
coordinate over the sphere $S^2\subset{R^3}$. We denote the
spherical coordinate of $S^{2}$ by $(\phi_1,\phi_2)$. A unit vector
$\vec{x}=(x_{1},x_{2},x_{3})$ can parameterized by
$x_1=\sin{\phi_1}\sin{\phi_2}$,
$x_2=\sin{\phi_1}\cos\phi_{2},~~~x_3=\cos{\phi_1}$. For any $0\leq
a\leq b\leq \frac{\pi}{2},$ we denote $\Omega_{a}^b=\{x\in S^{2}:
a\leq \phi_1(x)\leq b\}.$

{\bf{Theorem 1:}} For arbitrary two-qubit quantum state $\rho$ given
by (\ref{rho}), we have
\begin{eqnarray}\label{t1}
Q&\geq&\max
\left[\frac{4}{s_{ab}s_{cd}}|\int_{\Omega_a^b\times\Omega_c^d}<\vec{x},T\vec{y}>d\mu(\vec{x})d\mu(\vec{y})|
+\frac{2}{s^2_{cd}}\int_{\Omega_c^d\times\Omega_c^d}|T(\vec{x}-\vec{y})|d\mu(\vec{x})d\mu(\vec{y})\nonumber\right.\\
&&\left.+\frac{2}{s^2_{ab}}\int_{\Omega_a^b\times\Omega_a^b}|T^t(\vec{x}-\vec{y})|d\mu(\vec{x})d\mu(\vec{y})\right],
\end{eqnarray}
where $T^t$ stands for the transposition of $T$, and
$s_{\alpha\beta}=\int_{\Omega_{\alpha}^{\beta}}d\mu(\vec{x})$. The
maximum on the right side of the inequality goes over all the
integral area $\Omega_a^b\times\Omega_c^d$ with $0\leq a < b\leq
\frac{\pi}{2}$ and $0\leq c < d\leq \frac{\pi}{2}$.

See Methods for the proof of theorem 1.

The bound (\ref{t1}) can be calculated by parameterizing the
integral in terms of the sphere coordinates. Once a two-qubit is
given, the corresponding matrix $T$ is given. And the bound is
solely determined by $T$. This is similar to the CHSH inequality,
where the maximal violation is given by the two larger singular
values of $T$.

As an example, consider $T=diag(p_1,p_2,p_3)$, we have
\begin{eqnarray}
s_{ab}=\int_0^{2\pi}\int_a^b \sin \phi d\theta d\phi.
\end{eqnarray}
$s_{cd}$ in (\ref{t1}) are similarly given. The first two terms in
$s_{cd}$ (\ref{t1}) are given by
\begin{eqnarray}
&&\int_{\Omega_a^b\times\Omega_c^d}<\vec{x},T\vec{y}>d\mu(\vec{x})d\mu(\vec{y})
=\int_a^b\int_0^{2\pi}\int_c^d\int_0^{2\pi}f\sin\phi_1\sin\phi_2d\phi_1d\theta_1d\phi_2d\theta_2,\\
&&\int_{\Omega_a^b\times\Omega_c^d}|T(\vec{x}-\vec{y})|d\mu(\vec{x})d\mu(\vec{y})
=\int_a^b\int_0^{2\pi}\int_c^d\int_0^{2\pi}|g|\sin\phi_1\sin\phi_2d\phi_1d\theta_1d\phi_2d\theta_2,
\end{eqnarray}
where
$$
\ba{l} f=p_1\sin\phi_1\sin\theta_1\sin\phi_2\sin\theta_2
+p_2\sin\phi_1\cos\theta_1\sin\phi_2\cos\theta_2+p_3\cos\phi_1\cos\phi_2,\\
g=[p_1^2(\sin\phi_1\sin\theta_1-\sin\phi_2\sin\theta_2)^2
+p_2^2(\sin\phi_1\cos\theta_1-\sin\phi_2\cos\theta_2)^2+p_3^2(\cos\phi_1-\cos\phi_2)^2]^{\frac{1}{2}}.
\ea
$$
The last term in (\ref{t1}) is similarly to the second term, with
$T$ being replaced by $T^t$.

Thus for any given two-qubit quantum state, by substituting $T$ into
the integral, we have the lower bound of $Q$. The maximum taken over
$\Omega_a^b\times\Omega_c^d$ can be searched by varying the integral
ranges. The Werner state considered in
\cite{werner,barrett,062105,032112,hua} is a special case that
$p_1=p_2=p_3=p$. From our Theorem 1, we have that for $0.7054< x\leq
1$, the lower bound of $Q$ is always larger than that is derived
from the maximal violation of the CHSH inequality, see Fig. 1.

\begin{figure}[h]
\begin{center}
\resizebox{7cm}{!}{\includegraphics{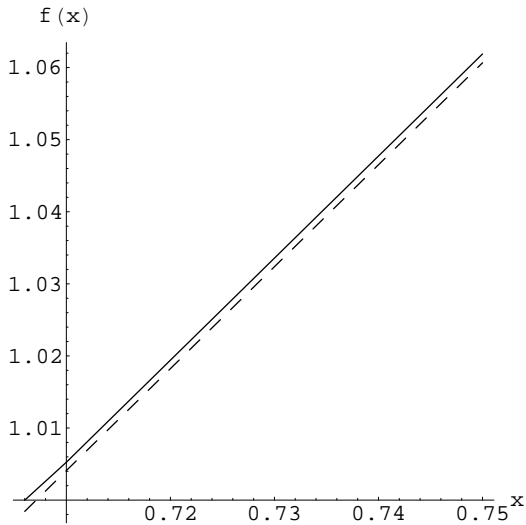}}
\end{center}
\caption{The lower bounds (denoted by $f(x)$) of $Q$ in Theorem 1
(solid line) and that obtained from the CHSH inequality (dashed
line). \label{fig1}}
\end{figure}

Let us now consider the generalized Bell diagonal two-qubit states
in detail, \be \rho_b=\frac{1}{4}(I\otimes
I-p_1\,\sigma_1\otimes\sigma_1-p_2\,\sigma_2\otimes\sigma_2-p_3\,\sigma_3\otimes\sigma_3).
\ee The positivity property requires that the parameters
$\{p_1,p_2,p_3\}$ must be inside a regular tetrahedron with vertexes
$\{-1,-1,1\},\{1,-1,-1\},\{1,1,1\},\{-1,1,-1\}$. By computing the
lower bound of $Q$ according to Theorem 1, we detect the regions
where the quantum states can never be described by LHV models, see
Fig. 2.

\begin{figure}[h]
\begin{center}
\resizebox{7cm}{!}{\includegraphics{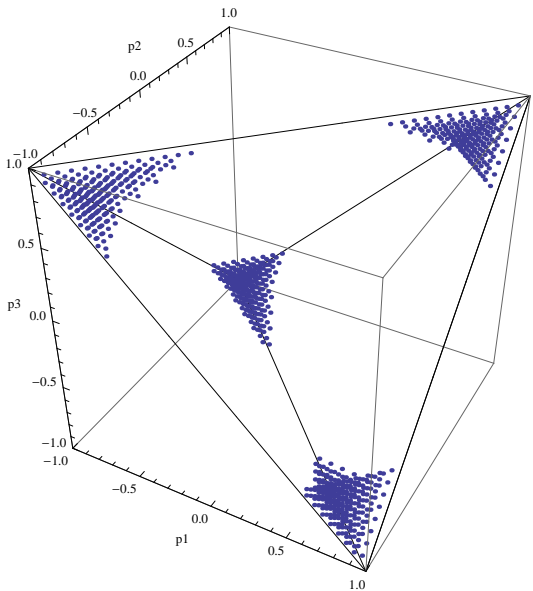}}
\end{center}
\caption{The quantum states $\rho_w$ that admits no LHV models are
listed by the points parameterized by $(p_1,p_2,p_3)$. \label{fig2}}
\end{figure}

By setting $p_1=0.9$, $p_2=0.9$ and $p_3=0.9$, one has the the
cross-sectional view, see Fig. 3.
\begin{figure}[h]
\begin{center}
\resizebox{7cm}{!}{\includegraphics{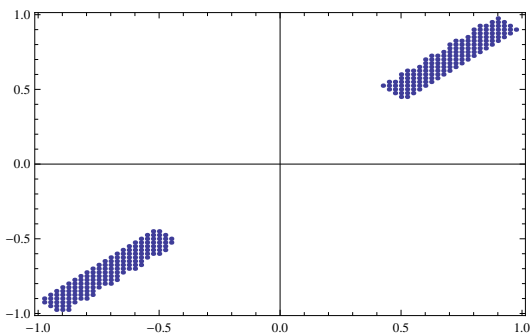}}
\end{center}
\caption{The same cross-sectional view of Fig. 2 for all p1 = 0.9,
p2 = 0.9 and p3 = 0.9. \label{fig3}}
\end{figure}

\medskip
{\sf High dimensional case}~ Generalizing our approach to high
dimensional case, now we study the nonlocality of general $d\times
d$ bipartite quantum states. To detect the nonlocality of a quantum
state, the important thing is to find a `good' Bell operator. For
even $d$, we set $\Gamma_1$, $\Gamma_2$ and $\Gamma_3$ to be
block-diagonal matrices, with each block an ordinary Pauli matrix,
$\sigma_1$, $\sigma_2$ and $\sigma_3$ respectively, as described in
\cite{vbell2} for $\Gamma_1$ and $\Gamma_3$. When $d$ is odd, we set
the elements of the $k$th row and the $k$th column in $\Gamma_1$,
$\Gamma_2$ and $\Gamma_3$ to be zero, with the rest elements of
$\Gamma_1$, $\Gamma_2$ and $\Gamma_3$ being the block-diagonal
matrices like the case of even $d$. Let $\Gamma_0$ be a $d\times d$
matrix whose only nonvanishing entry is $(\Gamma_0)_{mm}=1$ for
$m\in 1, 2, \cdots, d$, for odd $d$ and be a null matrix for even
$d$. We define observables $A=\vec{a}\cdot\vec{\Gamma}$ and
$B=\vec{b}\cdot\vec{\Gamma}$, where $\vec{\Gamma}=(\Gamma_0,
\Gamma_1, \Gamma_2, \Gamma_3)$, $\vec{a}=(1,a_1, a_2, a_3)$ and
$\vec{b}=(1, b_1, b_2, b_3)$ are vectors with norm $\sqrt{2}$. It is
easy to check that the eigenvalues of the observables $A$ and $B$
are either $1$ or $-1$.

We define the Bell operator to be \be\label{bo1}
I_d=\frac{1}{n^2}[\sum_{i,j=1}^{n}A_i\otimes B_j +\sum_{1\leq i <
j\leq n}C_{ij}\otimes(B_i-B_j)+\sum_{1\leq i < j\leq
n}(A_i-A_j)\otimes D_{ij}], \ee where $A_i, B_j, C_{ij}$ and
$D_{ij}$ are observables of the form $\vec{a}_i\cdot\vec{\Gamma},
\vec{b}_j\cdot\vec{\Gamma}, \vec{c}_{ij}\cdot\vec{\Gamma}$ and
$\vec{d}_{ij}\cdot\vec{\Gamma}$ respectively; $\vec{a}_{i},
\vec{b}_{j}, \vec{c}_{ij}$ and $\vec{d}_{ij}$ are vectors with norm
$\sqrt{2}$.

The Bell operator (\ref{bo1}) has the same structure as that in
(\ref{bo}), but fits for $d\times d$ quantum system. For a $d\times
d$ quantum state $\rho$, we set $\gamma$ to be a matrix with
elements $\gamma_{ij}=tr(\rho\Gamma_i\otimes \Gamma_j)$, $i,j=0, 1,
2, 3$. A lower bound of $Q$ defined in (\ref{q}) for $d\times d$
quantum system can be readily obtained as the follows.

{\bf{Theorem 2:}} For any quantum state $\rho$ in $d\times d$
quantum system ${\mathcal {H_{AB}}}$, we have that
\begin{eqnarray} Q&\geq&\max
\left[\left|\frac{1}{s_{ab}s_{cd}}\int_{\Omega_a^b\times\Omega_c^d}<\vec{x},\gamma\vec{y}>d\mu(\vec{x})d\mu(\vec{y})\right|\right.
\nonumber\\
&&\left.
+\frac{1}{2s^2_{cd}}\int_{\Omega_c^d\times\Omega_c^d}|\gamma(\vec{x}-\vec{y})|d\mu(\vec{x})d\mu(\vec{y})
+\frac{1}{2s^2_{ab}}\int_{\Omega_a^b\times\Omega_a^b}|\gamma^t(\vec{x}-\vec{y})|d\mu(\vec{x})d\mu(\vec{y})\right],
\end{eqnarray}
where $\gamma^t$ stands for the transposition of $\gamma$, and
$s_{\alpha\beta}=\int_{\Omega_{\alpha}^{\beta}}d\mu(\vec{x})$. The
maximum on the right side of the inequality is taken over all the
selection of integral area $\Omega_a^b\times\Omega_c^d$ with $0\leq
a < b\leq \frac{\pi}{2}$ and $0\leq c < d\leq \frac{\pi}{2}$.

See Methods for the proof of theorem 2.

According to the definition of $Q$ in (\ref{q}), we have that if the
lower bound for $Q$ in theorem 2 is larger than one, then a quantum
state in $d\times d$ bipartite quantum system can never be described
by an LHV model. The bound can readily calculated, similar to the
two-qubit case, once the matrix $\gamma$ for state is given.

Let us consider the isotropic state $\rho_I$ \cite{iso1,iso2}, a
mixture of the singlet state
$|\psi_+\ra=\frac{1}{\sqrt{3}}\sum_{i=1}^3|ii\ra$ and the white
noise: $\rho_I=\frac{1-x}{d^2}I+x|\psi_+\ra\la\psi_+|$, $0\leq x\leq
1$. $\rho_I$ is entangled for $x>\frac{1}{8}(-1+\frac{9}{d})$. For
$d=3$, $\rho_I$ is entangled for $x>1/4$. From Theorem 2, $\rho_I$
is nonlocal for $x>0.7653$.

As another example we consider the state $\rho$ from mixing the
singlet state $|\psi_+\ra$ with
$\sigma=\frac{1}{4}(I_3-\Gamma_0)\otimes(I_3-\Gamma_0)-\frac{\alpha}{4}\sum_{i=2}^4\Gamma_i\otimes\Gamma_i$,
$\rho=(1-\beta)\sigma+\beta|\psi_+\ra\la\psi_+|$. One can list by
Theorem 2 the points that admit no LHV model, see Fig. 4.

\begin{figure}[h]
\begin{center}
\resizebox{7cm}{!}{\includegraphics{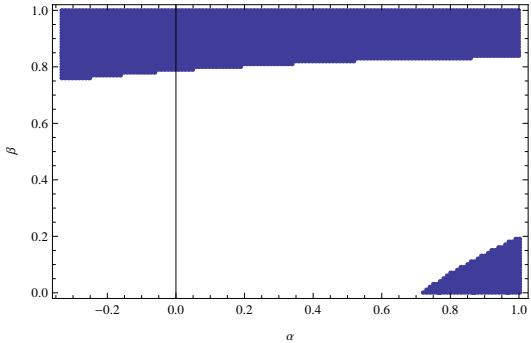}}
\end{center}
\caption{(Color on line) Quantum states $\rho$ parameterized by
$(\alpha, \beta)$ that admit no LHV model (blue regions).
\label{fig3}}
\end{figure}

\medskip
\noindent{\bf Discussions}

Nowadays, quantum nonlocality is a fundamental subject in quantum
information theory such as quantum cryptography, complexity theory,
communication complexity, estimates for the dimension of the
underlying Hilbert space, entangled games, etc\cite{jung}. Thus it
is a basic
 question to check and to qualify the nonlocality of
a quantum state. We have derived an analytical and computable lower
bound of the quantum violation by using a Bell inequality with
infinitely many measurement settings. The bound is shown to be
better than that is obtained from the CHSH inequality and the
discrete models. Sufficient conditions for the LHV description of
high dimensional quantum states have also derived. Apart from the
computation of maximal violations for bipartite Bell inequalities,
our methods can also contribute to the analysis of the nonlocality
of multipartite quantum systems.

\medskip
\noindent{\bf Methods}
\medskip

{\sf Proof of Theorem 1}~ For any two-qubit quantum state $\rho$
given in (\ref{rho}), we have
\begin{eqnarray}
Q&\geq&\max|\la I \ra|=\max\frac{1}{n^2}\left|\sum_{i,j=1}^n
tr(A_i\otimes B_j\,\rho)+\sum_{1\leq i<j\leq n}
tr(C_{ij}\otimes(B_i-B_j)\,\rho)\right.\nonumber\\
&&\quad\quad\quad\quad\quad\quad\left.+\sum_{1\leq i<j\leq n}
tr((A_i-A_j)\otimes D_{ij}\,\rho)\right|\nonumber\\
&=&\max\frac{4}{n^2}\left|\sum_{i,j=1}^n\sum_{k,l=1}^3
a_{ik}b_{jl}t_{kl} +\sum_{1\leq i<j\leq n}\sum_{k,l=1}^3
c_{ij,k}(b_{il}-b_{jl})t_{kl}
+\sum_{1\leq i<j\leq n}\sum_{k,l=1}^3 (a_{ik}-a_{jk})d_{ij,l}t_{kl}\right|\nonumber\\
&=&\max\frac{4}{n^2}\left|\sum_{i,j=1}^n\la\vec{a}_i,T\vec{b}_j\ra
+\sum_{1\leq i<j\leq n}\la\vec{c}_{ij},T(\vec{b}_i-\vec{b}_j)\ra+\sum_{1\leq i<j\leq n}\la T^t(\vec{a}_{i}-\vec{a}_j),\vec{d}_{ij}\ra\right|\nonumber\\
&=&\max\frac{4}{n^2}\left[|\sum_{i,j=1}^n\la\vec{a}_i,T\vec{b}_j\ra|
+\sum_{1\leq i<j\leq n}|T(\vec{b}_i-\vec{b}_j)|+\sum_{1\leq i<j\leq
n}|T^t(\vec{a}_{i}-\vec{a}_j)|\right].
\end{eqnarray}
Under the limit $n\to\infty$, we have
\begin{eqnarray}
Q&\geq&\max\left[\frac{4}{s_{ab}s_{cd}}|\int_{\Omega_a^b\times\Omega_c^d}<\vec{x},T\vec{y}>d\mu(\vec{x})d\mu(\vec{y})|
+\frac{2}{s^2_{cd}}\int_{\Omega_c^d\times\Omega_c^d}|T(\vec{x}-\vec{y})|d\mu(\vec{x})d\mu(\vec{y})\right.\nonumber\\
&&\left.+\frac{2}{s^2_{ab}}\int_{\Omega_a^b\times\Omega_a^b}|T^t(\vec{x}-\vec{y})|d\mu(\vec{x})d\mu(\vec{y})\right],
\end{eqnarray}
which proves (\ref{t1}). \hfill \rule{1ex}{1ex}

{\sf Proof of Theorem 2}~ With the special selected observables of
the form $\vec{a}\cdot \Gamma$ for $d\times d$ quantum systems, we
have that
\begin{eqnarray}
Q&\geq&\max|\la I_d \ra|=\max|\frac{1}{n^2}[\sum_{i,j=1}^n
tr(A_i\otimes B_j\rho)+\sum_{1\leq i<j\leq n}
tr(C_{ij}\otimes(B_i-B_j)\rho)\nonumber\\
&&\quad\quad\quad\quad\quad\quad+\sum_{1\leq i<j\leq n}
tr((A_i-A_j)\otimes
D_{ij}\rho)]|\nonumber\\
&=&\frac{1}{n^2}\max|[\sum_{i,j=1}^n\sum_{k,l=0}^3
a_{ik}b_{jl}\gamma_{kl}+\sum_{1\leq i<j\leq n}\sum_{k,l=0}^3
(c_{ij,k}(b_{il}-b_{jl})\gamma_{kl}+(a_{ik}-a_{jk})d_{ij,l}\gamma_{kl})]|\nonumber\\
&=&\frac{1}{n^2}\max|[\sum_{i,j=1}^n\la\vec{a}_i,\gamma\vec{b}_j\ra
+\sum_{1\leq i<j\leq n}(|\gamma(\vec{b}_i-\vec{b}_j)|+|
\gamma^t(\vec{a}_{i}-\vec{a}_j)|]|\nonumber\\
&\geq&\max
[|\frac{1}{s_{ab}s_{cd}}\int_{\Omega_a^b\times\Omega_c^d}<\vec{x},\gamma\vec{y}>d\mu(\vec{x})d\mu(\vec{y})|
\nonumber\\
&&+\frac{1}{2s^2_{cd}}\int_{\Omega_c^d\times\Omega_c^d}|\gamma(\vec{x}-\vec{y})|d\mu(\vec{x})d\mu(\vec{y})
+\frac{1}{2s^2_{ab}}\int_{\Omega_a^b\times\Omega_a^b}|\gamma^t(\vec{x}-\vec{y})|d\mu(\vec{x})d\mu(\vec{y})],
\end{eqnarray}
where in the last step, we have taken the limit $n \to \infty$.
\hfill \rule{1ex}{1ex}

\newpage
\bigskip
\noindent{\sf Acknowledgements}

\noindent This work is supported by the NSFC 11105226, 11275131; the
Fundamental Research Funds for the Central Universities
No.12CX04079A, No.24720122013; Research Award Fund for outstanding
young scientists of Shandong Province No.BS2012DX045.

\bigskip
\noindent{\sf Author contributions}

\noindent  M. Li and S.M. Fei wrote the main manuscript text. T.
Zhang, B. Hua, and X.Q. Li-Jost computed the examples. All authors
reviewed the manuscript.

\bigskip
\noindent{\sf Additional Information}

\noindent Competing Financial Interests: The authors declare no competing financial interests.

\end{document}